**Genomics-guided molecular maps of coronavirus targets in human cells: a path toward the repurposing of existing drugs to mitigate the pandemic**

Gennadi V. Glinsky[1]

[1] Institute of Engineering in Medicine

University of California, San Diego

9500 Gilman Dr. MC 0435

La Jolla, CA 92093-0435, USA

Correspondence: gglinskii@ucsd.edu

Web: http://iem.ucsd.edu/people/profiles/guennadi-v-glinskii.html

**Running title:** Genomics-guided mitigation maps for coronavirus pandemic

**Abstract**

Human genes required for SARS-CoV-2 entry into human cells, *ACE2* and *FURIN*, were employed as baits to build genomics-guided maps of up-stream regulatory elements, their expression and functions in human body, including pathophysiologically-relevant cell types. Genes acting as repressors and activators of the *ACE2* and *FURIN* genes were identified based on the analyses of gene silencing and overexpression experiments as well as relevant transgenic mouse models. Panels of repressors (*VDR; GATA5; SFTPC; HIF1a*) and activators (*HMGA2; INSIG1*) were then employed to identify existing drugs that could be repurposed to mitigate the coronavirus infection. Present analyses identify Vitamin D and Quercetin as promising pandemic mitigation agents. Gene expression profiles of Vitamin D and Quercetin activities and their established safety records as over-the-counter medicinal substances suggest that they may represent viable candidates for further assessment and considerations of their potential as coronavirus pandemic mitigation agents. Notably, gene set enrichment analyses and expression profiling experiments identify multiple drugs, most notably testosterone, dexamethasone, and doxorubicin, smoking, and many disease conditions that appear to act as putative coronavirus infection-promoting agents. Discordant patterns of Testosterone versus Estradiol impacts on SCARS-CoV-2 targets suggest a plausible molecular explanation of the apparently higher male mortality during coronavirus pandemic.

**Introduction**

Coronavirus pandemic 2020 caused by the newly emerged SARS-CoV-2 virus is rapidly entering the most dangerous acute phase of its evolution in the United States. Absence of the vaccine and lack of efficient targeted therapeutic approaches emphasizes the urgent need for identification of candidate pandemic mitigation agents among existing drugs and medicinal substances.

SARS-CoV-2 virus was discovered in December 2019 and shortly thereafter it was isolated and sequenced (Zhou et al., 2020; Zhu et al., 2020). Recent analyses of the structure, function, and antigenicity of the SARS-CoV-2 spike glycoprotein revealed the key role of the *ACE2* and *FURIN* genes in facilitating the high-affinity binding of viral particles and their entry into human cells (Walls et al., 2020). The efficient invasion of host cells by the SARS-CoV-2 is further enhanced by the presence of the unexpected furin cleavage site, which is cleaved during the biosynthesis (Walls et al., 2020). This novel feature distinguishes the previously known SARS-CoV and the newly emerged SARS-CoV-2 viruses and possibly contributes to the expansion of the cellular tropism of the SARS-CoV-2 (Walls et al., 2020). Collectively, these observations identified protein products of the human genes *ACE2* and *FURIN* as the high-affinity receptor (ACE2) and invasion-promoting protease (FURIN) acting as the principal mediators of the SARS-CoV-2 invasion into human cells.

In this contribution, genomic screens were performed employing the *ACE2* and *FURIN* genes as baits to build genomics-guided human tissues-tailored maps of up-stream regulatory elements, their expression and functions. To identify the high-priority list of potential candidate mitigation agents, the validation analyses were performed using gene silencing and overexpression experiments as well as relevant transgenic mouse models with the emphasis on pathophysiologically-relevant cell types. Panels of repressors (*VDR; GATA5; SFTPC*) and activators (*HMGA2; INSIG1*) of the *ACE2* and *FURIN* expression were identified and then employed to identify existing drugs and medicinal substances that could be repurposed to ameliorate the outcomes of the coronavirus infection. Two of the most promising candidate mitigation agents, namely Vitamin D and Quercetin, manifest gene expression-altering activities and have established safety records as over-the-counter medicinal substances that seem sufficient for further assessment and considerations of their potential utility for amelioration of the clinical course of coronavirus pandemic. Unexpectedly, present analyses revealed discordant patterns of Testosterone versus Estradiol impacts on SCARS-CoV-2 targets with the former manifesting the potential coronavirus infection-promoting activities, which is consistent with the apparently higher male mortality across all age groups during the coronavirus pandemic.

**Results**

**Gene set enrichment analyses (GSEA) of genomic features associated with the *ACE2* and *FURIN* genes**

One of the goals of this work was to identify human genes implicated in regulatory cross-talks affecting expression and functions of the *ACE2* and *FURIN* genes to build a model of genomic regulatory interactions potentially affecting the SCARS-CoV-2 coronavirus infection. To this end, GSEA were carried out using the *ACE2* and *FURIN* genes as baits applied to a broad spectrum of genomic databases reflecting the current state of knowledge regarding the structural, functional, regulatory, and pathophysiological features that could be statistically linked to these genes. Expression profiling experiments and GSEA revealed ubiquitous patterns of both *ACE2* and *FURIN* genes across human tissues (Supplemental Figure S1) with notable examples of high expression of the *FURIN* gene in the lung (second-ranked tissue in the GTEX database) and testis being identified as the top-ranked *ACE2*-expressing tissue. In addition to the human lung tagged by the *ACE2* expression in the ACRHS4 Human Tissues database search, other noteworthy significantly enriched records are the Peripheral Blood Mononuclear Cells (PBMC),

Natural Killer Cells and Macrophages tagged by the *FURIN* expression (Supplemental Figure S1).

GSEA of the virus perturbations' data sets among Gene Expression Omnibus (GEO) records of up-regulated genes identified the SARS-CoV challenge at 96 hrs (GSE47960) as the most significantly enriched record (Supplemental Figure S2) tagged by expression of both *ACE2* and *FURIN* in human airway epithelial cells. These observations suggest that coronavirus infection triggers the increased expression of both *ACE2* and *FURIN* genes 4 days after the initial encounter with host cells (Figure 1; Supplemental Figure S2). These findings were corroborated by the increased *FURIN* expression documented in the PBMC of patients with severe acute respiratory syndrome (Figure 1; Supplemental Figure S1; Reghunathan et al., 2005). It would be of interest to investigate whether this potentially infection-promoting effect on expression of the host genes in virus-targeted cells is mediated by the virus-induced release of the biologically-active molecules with the paracrine mode of actions such as interleukins and cytokines.

GSEA identified numerous significantly enriched records of common human disorders manifesting up-regulation of either *ACE* or *FURIN* genes (Supplemental Figure S3), which is consistent with the clinical observations that individuals with underlying health conditions are more likely to have clinically severe and lethal coronavirus infection. Similarly, exploration of the DisGeNET database of human disorders highlighted multiple disease states' records manifesting altered expression of either *ACE2* of *FURIN* genes (Supplemental Figure S3). Cigarette smoking appears to significantly increase the *ACE2* expression in human large airway epithelial cells (Supplemental Figure S3), indicating that cigarette smoking should be considered as a potential coronavirus infection-promoting agent.

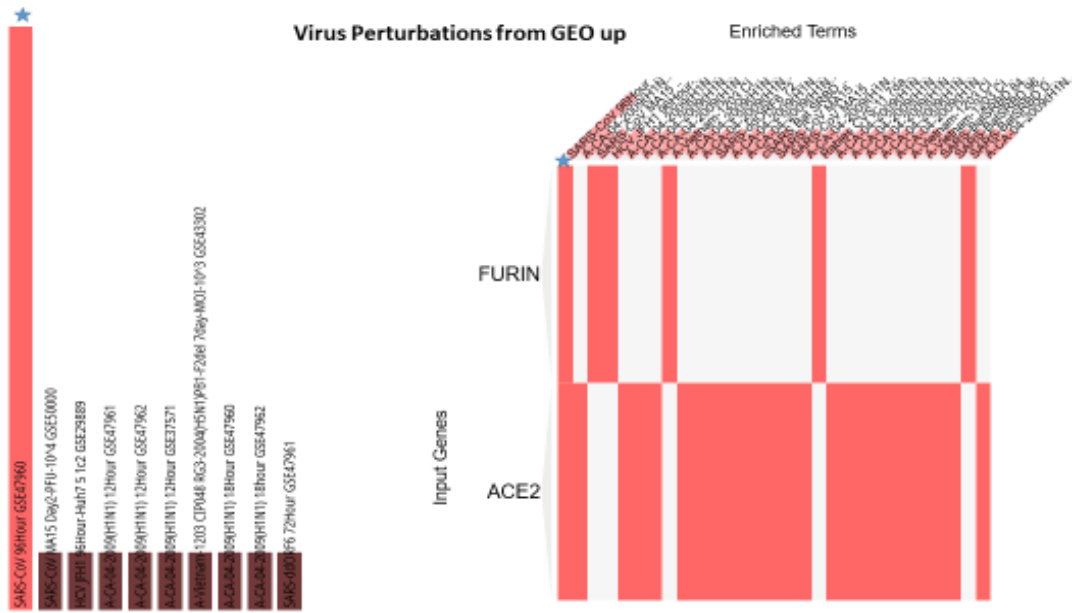

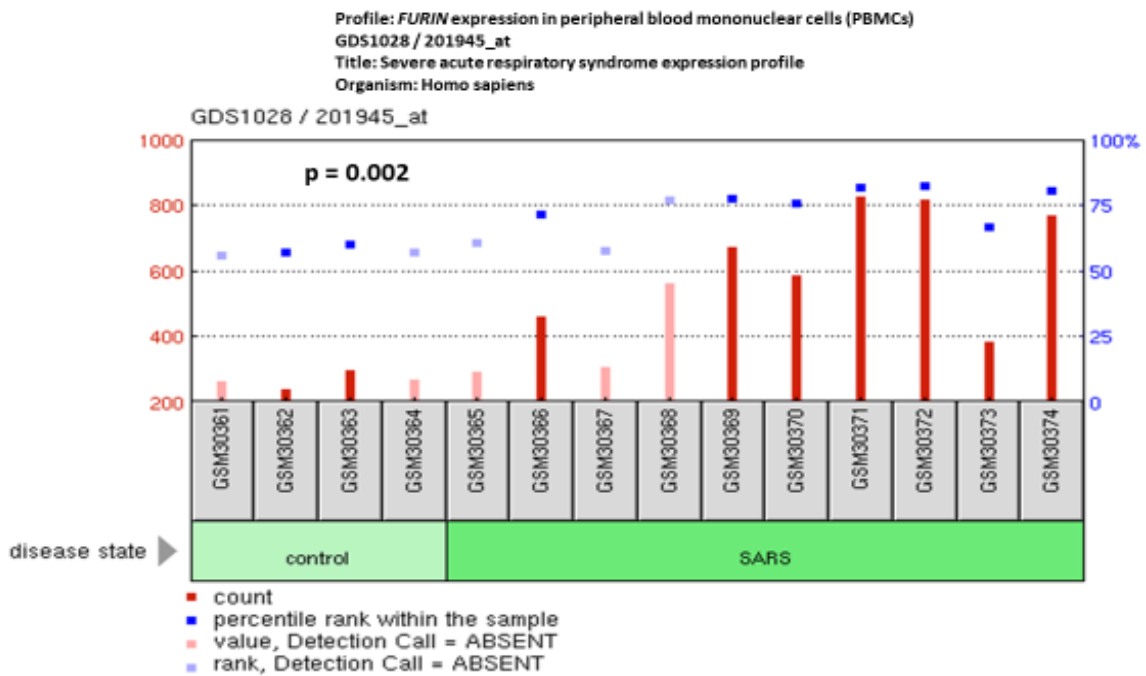

**Figure 1. Effects of viral challenges on expression of the *ACE2* and *FURIN* genes.**

  a. Gene Set Enrichment analyses of the Virus Perturbations from GEO focused on up-regulated genes.

  b. Increased *FURIN* expression in peripheral blood mononuclear cells (PBMC) of patients with Severe Acute Respiratory Syndrome (SARS).

Gene Ontology (GO) analyses revealed that *ACE2* and *FURIN* genes are associated with the largely non-overlapping records of GO Biological Processes, GO Molecular Functions, and GO Cellular Components (Supplemental Figure S4). The common significantly enriched records are Viral Life Cycle (GO Biological Process 2018); Peptidase activity (acting on L-amino acid peptides) and Endopeptidase activity (GO Molecular Function 2018); Membrane raft (GO Cellular Component 2018); Meprin A complex and Retrotrasposon nucleocapsid (Jensen Compartments).

**Identifications of the enriched records of transcription factor-binding sites affecting the *ACE2* and *FURIN* expression**

GSEA of the enriched records of transcription factors' binding sites (TFBS) using ENCODE TF ChIP-seq 2015 and ChEA 2016 databases revealed predominantly distinct patterns of TFBS associated with the *ACE2* and *FURIN* genes (Supplemental Figure S5). Common TFBS shared by both *ACE2* and *FURIN* genes are *FOS, JUND, EP300* (ENCODE TF ChIP-seq 2015) and *GATA1, GATA2, RUNX1, FOXA1, HNF4A* (ChIP-seq 2015). Consistent with these findings, non-overlapping profiles of significantly enriched records associated with either *ACE2* or *FURIN* genes were observed of pathways (BioPlanet 2019 database), protein-protein interactions (PPI) hub proteins (PPI Hub Proteins database), and drugs affecting *ACE2* and *FURIN* expression (Drug Signatures Database, DSigDB), indicating that regulatory mechanisms governing the expression and activities of the *ACE2* and *FURIN* genes are predominantly discordant (Supplemental Figure S5).

Next, the Gene Expression Omnibus (GEO) database was interrogated to gauge the effects on *ACE2* and *FURIN* expression of transcription factors having TFBS associated with their promoters. There are multiple relevant GEO records reporting the activation effects of the *JNK1/c-FOS* pathway on *ACE2* and *FURIN* expression as well as the activation effects of *FURIN* depletion on expression of the *Fos, Jun, Jund,* and *Junb* genes (Supplemental Figure

S5). Conversely, *c-Jun* inhibition (effect of the dominant negative *c-Jun*) or *c-Jun* depletion (*c-Jun* knockout) has resulted in decreased expression of the *FURIN* gene (Supplemental Figure S5). The summary of these observations is reported in the Figure 2.

Similarly, there are several reports indicating that depletion of either *Hnf4a* or *Runx1* in mouse cells and *RUNX1* in human cells decreases the *ACE2* and *FURIN* expression (Supplemental Figure S6). Conversely, *FURIN* depletion enhances expression of the *Runx1* and *Foxa1* genes in murine T cells (Supplemental Figure S6). In contrast, *FURIN* depletion decreases expression of the *Hnf4a* gene, while *Hnf4a* depletion decreases the *FURIN* gene expression (Supplemental Figure S6). The summary of these observations is reported in the Figure 2.

**Identification of the *VDR* and *HIF1a* genes as putative repressors of the *ACE2* expression**

Next GSEA of genomic databases were performed to identify the potential activators and repressors of the *ACE2* and *FURIN* genes. Analysis of the ARCHS4 transcription factors' co-expression database identified the *VDR* genes that co-expressed with both *ACE2* and *FURIN* genes in human tissues (Supplemental Figure S7). Other significantly enriched records manifest non-overlapping patterns of co-expression with either *ACE2* of *FURIN* genes. The GTEX expression profile of the *VDR* gene in human tissues revealed the ubiquitous pattern of expression and placed the *VDR* expression in human lungs in the top quartile (Supplemental Figure S7). Analysis of gene expression profiling experiments of wild type and vitamin D receptor (Vdr) knockout primary bone marrow-derived macrophages reported by Helming et al. (2005) demonstrate increased expression of the *ACE2* gene in the *VDR* knockout cells (Supplemental Figure S7) implicating the product of the *VDR* gene as the putative repressor of the *ACE2* expression. Consistent with this hypothesis, Vitamin D appears to inhibit the *ACE2* expression in human bronchial smooth muscle cells (Supplemental Figure S7).

A

JNK1/c-FOS pathway-associated activation of the *ACE2* and *FURIN* expression may trigger the auto-regulatory negative feed-back loop of the FURIN-mediated repression of the expression of *JUN, JUNB, JUND,* and *c-FOS* genes

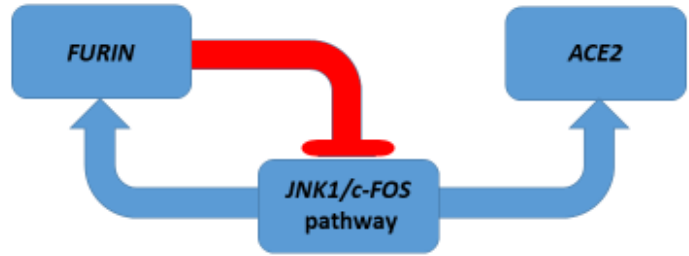

B

*RUNX1* pathway-associated activation of the *ACE2* and *FURIN* expression may trigger the auto-regulatory negative feed-back loop of the *FURIN*-mediated repression of the *RUNX1* gene expression

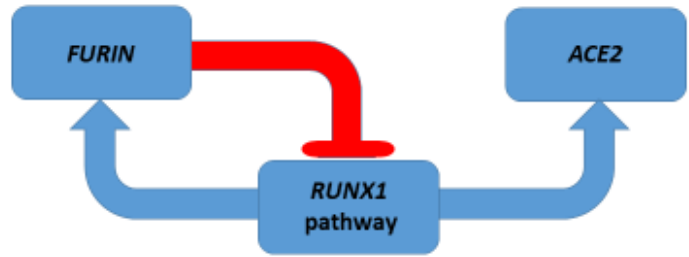

c

HNF4a pathway-associated activation of the ACE2 and FURIN expression may trigger the auto-regulatory positive feed-back loop of the FURIN-mediated activation of the HNF4a expression

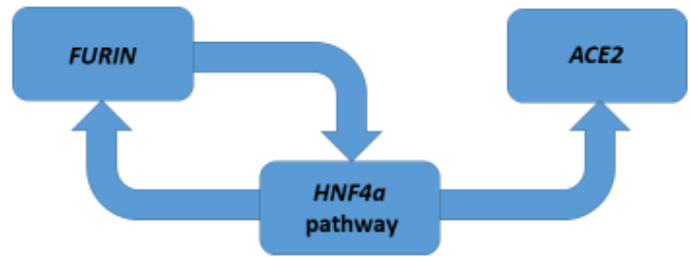

**Figure 2. Pathways and genes affecting the newly emerged SARS-CoV-2 virus-related host targets.**

a. *JNK1/c-FOS* pathway-associated activation of the *ACE2* and *FURIN* expression may trigger the auto-regulatory negative feed-back loop of the FURIN-mediated repression of the expression of *JUN, JUNB, JUND,* and *c-FOS* genes.

b. *RUNX1* pathway-associated activation of the *ACE2* and *FURIN* expression may trigger the auto-regulatory negative feed-back loop of the *FURIN*-mediated repression of the *RUNX1* gene expression

c. *HNF4a* pathway-associated activation of the *ACE2* and *FURIN* expression may trigger the auto-regulatory positive feed-back loop of the *FURIN*-mediated activation of the *HNF4a* expression.

Notably, examinations of direct and reciprocal effects of the *VDR* gene and Vitamin D administration on expression of the *JNK1/c-FOS* pathway genes revealed the expression profiles consistent with the potential therapeutic utility of the Vitamin D administration and activation of the *VDR* gene expression (Supplemental Figure S7). Analyses of direct and reciprocal effects of the *VDR* gene and Vitamin D administration on the *HNF4a* expression revealed that *HNF4a* depletion in human and murine cells inhibits the *VDR* gene expression, while the *Vdr* gene depletion increases the *Hnf4a* expression (Supplemental Figure S7). These result are consistent with the hypothesis stating that Vitamin D administration and activation of the *VDR* gene expression may have mitigating effects on the coronavirus infection. The summary of these findings is reported in the Figure 3.

GSEA of the Transcription Factor's Perturbations Followed by Expression database and GEO Gene Perturbations database focused on up-regulated genes identified *HIF1a* and *POU5F1* gene products as putative repressors of the *ACE2* and *FURIN* expression (Supplemental Figure S8). These findings were corroborated by observations that *HIF1a* overexpression in human embryonic kidney cells significantly inhibits the *ACE2* expression (Supplemental Figure S8). Notably, Vitamin D significantly increases expression of the *HIF1a* gene in human bronchial smooth muscle cells (Supplemental Figure S8), suggesting that *VDR* and *HIF1A* genes may cooperate as repressors of the *ACE2* expression.

**GSEA identify Estradiol and Quercetin as putative candidate coronavirus infection mitigation agents.**

GSEA of the Drug Perturbations from GEO database focused on down-regulated genes identified Estradiol and Quercetin among the top significantly enriched records (Supplemental Figure S9). Estradiol appears to affect both *FURIN* and *ACE2* expression, while Quercetin seems to target the *ACE2* expression. Consistently, GSEA of the Ligand Perturbations from GEO focused on down-regulated genes identified five of Estradiol administration records (50%) among top ten significantly enriched ligand perturbations records (Supplemental Figure S9).

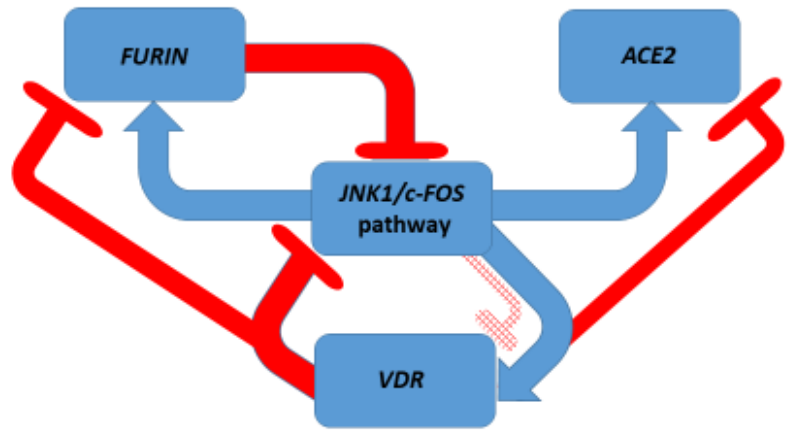

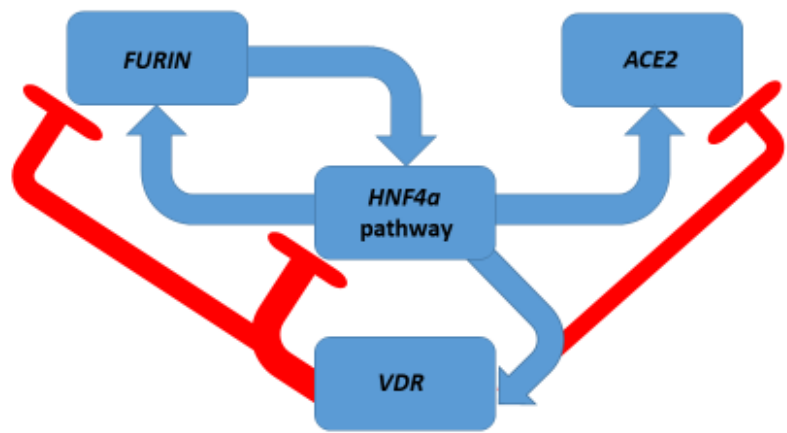

**Figure 3. Effects of the *VDR* gene and Vitamin D on pathways and genes affecting the newly emerged SARS-CoV-2 virus-related host targets.**

   a.  *JNK1/c-FOS* pathway-associated activation of the *ACE2* and *FURIN* expression may trigger the auto-regulatory negative feed-back loop of the FURIN-mediated repression of the expression of *JUN, JUNB, JUND,* and *c-FOS* genes.

   b.  *HNF4a* pathway-associated activation of the *ACE2* and *FURIN* expression may trigger the auto-regulatory positive feed-back loop of the *FURIN*-mediated activation of the *HNF4a* expression.

GSEA of the Drug Perturbations from GEO database focused on up-regulated genes indicated that doxorubicin, imatinib, and bleomycin may act as potential coronavirus infection-promoting agents (data not shown). Collectively, these observations provide the initial evidence supporting the hypothesis that both Estradiol and Quercetin may function as potential candidate coronavirus infection mitigation agents.

Consistent with this hypothesis, interrogation of the GEO records revealed that Quercetin appears to inhibit expression of several potential coronavirus infection-promoting genes: *c-FOS* expression in human and rat cells (Supplemental Figure S9); *Runx1* expression in rat cells (Supplemental Figure S9); *HNF4a* expression in human cells (Supplemental Figure S9). However, Quercetin administration appears to increase *c-Fos* expression in cultured rat cardiomyocytes (Supplementary Figure S9).

**Confirmation of the Estradiol and Quercetin activities as potential candidate coronavirus infection mitigation agents.**

Results of GSEA suggest that both Estradiol and Quercetin appear to exhibit biological activities consistent with the activity of medicinal compounds expected to mitigate the coronavirus infection. Next, manual curation of the GEO data sets has been carried out to identify further experimental evidence supporting this hypothesis. Administration of Estradiol appears to inhibit *ACE2* and/or *FURIN* expression in rat, mouse, and human cells (Supplemental Figure S10) and the effects of Estradiol seem to be mediated by the estrogen receptor beta. In agreement with the hypothesis on potential therapeutic utility of the Quercetin, administration of Quercetin has resulted in significantly decreased expression of the *ACE2* gene during differentiation of human intestinal cells (Supplemental Figure S10).

However, Estradiol administration appears to manifest the cell type-specific effects on c-FOS expression (Supplemental Figure S10). For example, it decreases the c-FOS expression in endometrium of Macaca mulatta while it increases c-FOS expression in the mouse uterus (Supplemental Figure S10). These observations indicate that any definitive conclusions regarding the potential clinical utility of identified herein potential coronavirus infection mitigating agents should be made only after appropriately designed and carefully executed preclinical studies and randomized clinical trials. In contrast to the Estradiol, which exhibit evidence of both putative coronavirus infection-mitigating actions and coronavirus infection-promoting activities, administration of Testosterone appears to manifest more clearly-defined patterns of altered gene expression consistent with Testosterone being identified as the potential coronavirus infection-promoting agent (Supplemental Figure S11).

**Potential mechanisms affecting gene expression inferred from transgenic mouse models and observed in pathophysiologically and therapeutically relevant mouse and human cells.**

Taken into considerations that the effects of potential coronavirus infection mitigation agents often manifest cell type-specific patterns of gene expression changes, next the manual curation of the GEO gene expression profiles were carried out to identify the relevant host genetic targets and putative mitigation agents. These analyses identified several candidate repressors (*VDR; GATA5; SFTC; HIF1a*) and activators (*INSIG1; HMGA2*) of the *ACE2* and *FURIN* expression (Supplemental Figure S12). Notably, the effects on gene expression of the administration of either Vitamin D or Quercetin appear consistent with their definition as putative coronavirus infection mitigation agents (Supplemental Figure S12). The summary of these observations is presented in the Figure 4. The conclusion regarding the findings of cell type-specific effects on gene expression of putative coronavirus infection mitigating agents remains valid and examples of the potential negative effects of drugs on the ACE2 expression are reported in the Supplemental Figure S12). For example, the *HIF1a* expression is significantly increased in murine alveolar type I cells deficient in sterol-response element-binding proteins inhibitor Insig1 (Supplemental Figure S6). These data indicate that the *INSIG1* gene product, which appears to function as activator of the *ACE2* expression, may function as the inhibitor of the *HIF1a* expression, thus interfering with the *HIF1a*-mediated *ACE2* repression in specific cell types. Additional examples of the potential positive and negative effects on gene expression inferred from transgenic mouse models are reported in the Supplemental Figure S13.

**Conclusion**

The main motivation of this work was to identify human genes implicated in regulatory cross-talks affecting expression and functions of the *ACE2* and *FURIN* genes to build a model of genomic regulatory interactions potentially affecting the SCARS-CoV-2 coronavirus infection. A panel of genes acting as activators and/or repressor of the *ACE2* and/or *FURIN* expression then could be employed to search for existing drugs and medicinal substances that, based on their mechanisms of activities, could be defined as the candidate coronavirus infection mitigation agents. After experimental and clinical validation, these existing drugs could be utilized to ameliorate the clinical severity of the pandemic.

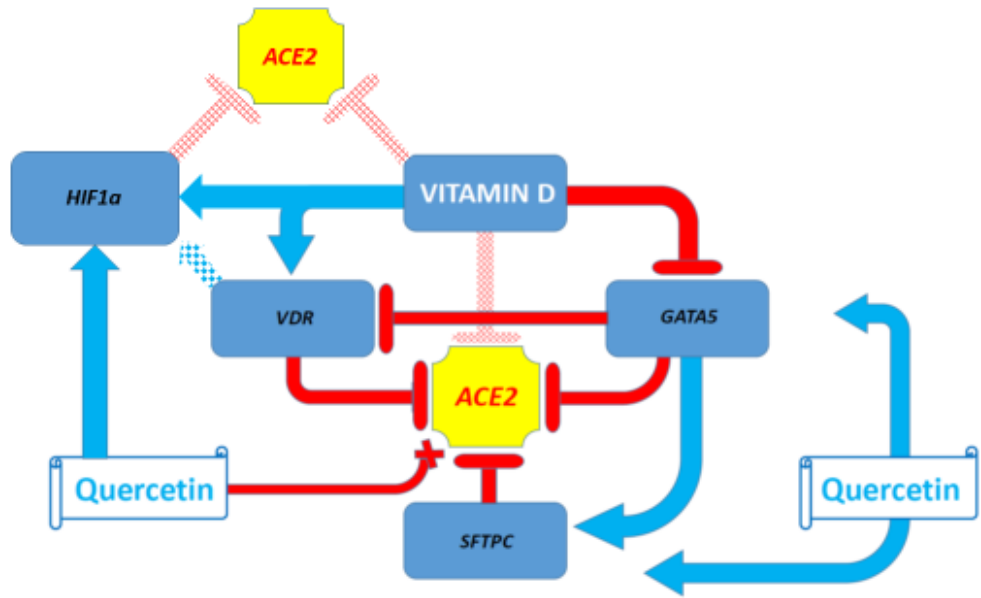

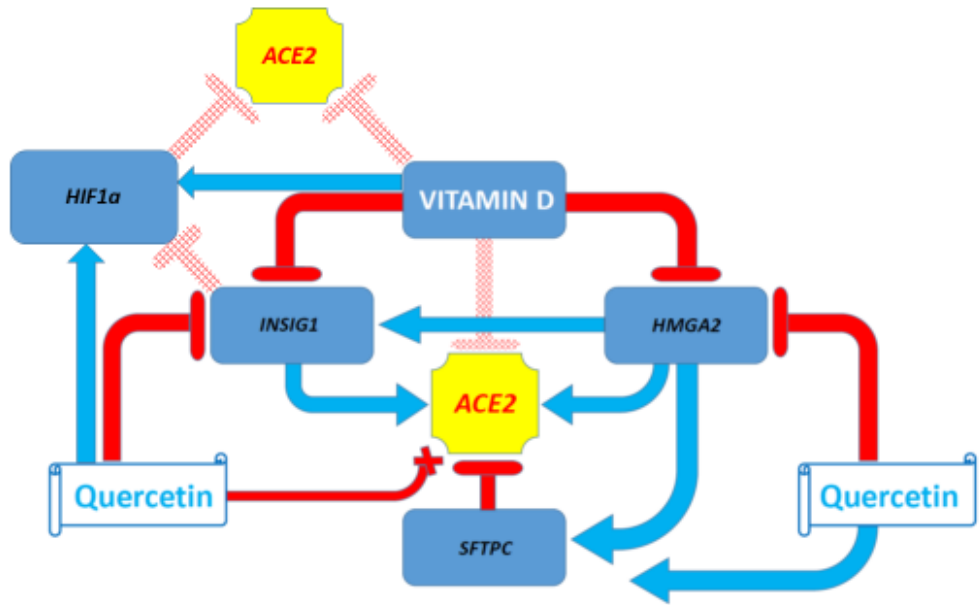

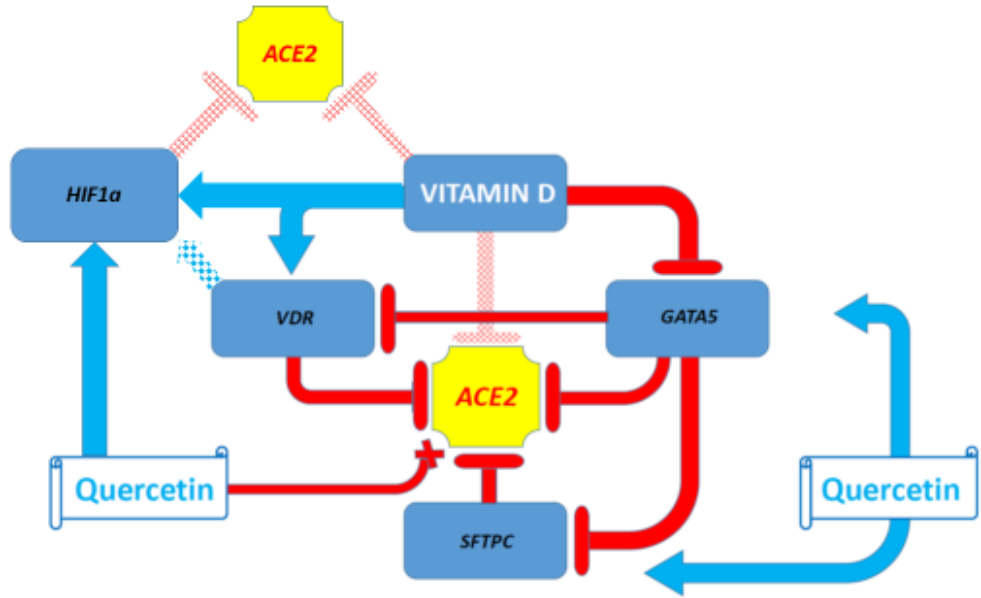

GATA5 inhibits SFTPC expression in the mouse lungs

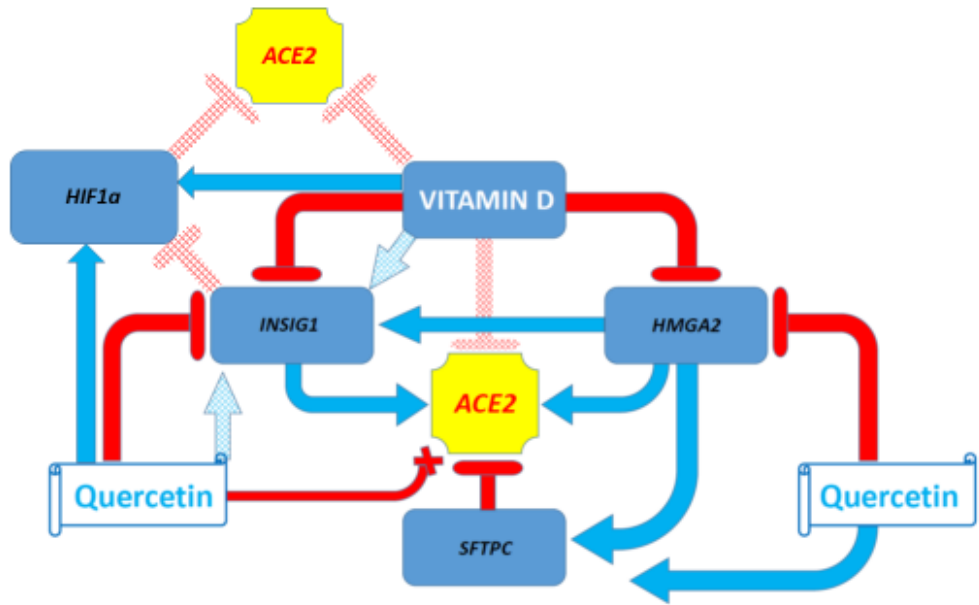

Vitamin D enhances *INSIG1* expression in human bronchial smooth muscle cells

Quercetin enhances *INSIG1* expression in human intestinal cells

**Figure 4. Effects of the *VDR* gene, Vitamin D, and Quercetin on pathways and genes affecting the newly emerged SARS-CoV-2 virus-related host targets.**

  a. Effects of the *VDR* gene, Vitamin D, and Quercetin on repressors of the *ACE* expression.
  b. Effects of the *VDR* gene, Vitamin D, and Quercetin on activators of the *ACE* expression.

c. Effects of the *VDR* gene, Vitamin D, and Quercetin on repressors of the *ACE* expression reflecting *GATA5* inhibitory effects on *SFTPC* expression in the mouse lungs.
   d. Effects of the *VDR* gene, Vitamin D, and Quercetin on activators of the ACE expression reflecting the cell type-specific effects of Vitamin D and Quercetin: Vitamin D-induced activation of the *INSIG1* expression in human bronchial smooth muscle cells and Quercetin-induced activation of the *INSIG1* expression in human intestinal cells.

This knowledge could also be exploited in an ongoing effort to discover novel targeted therapeutics tailored to block the SCARS-CoV-2 infection.

One of the important findings documented herein is that identified medicinal compounds with potential coronavirus infection-mitigating effects also appear to induce cell type-specific patterns of gene expression alterations. Therefore, based on all observations reported in this contribution, it has been concluded that any definitive recommendations regarding the potential clinical utility of identified herein potential coronavirus infection mitigating agents, namely Vitamin D and Quercetin, should be made only after appropriately designed and carefully executed preclinical studies and randomized clinical trials.

Present analyses highlighted the major uncertainty regarding the outcomes of the current pandemic associated with the potential of the SCARS-CoV-2 virus for the expansion of the cellular tropism (Walls et al., 2020) based on access to genetically vulnerable host cells due to nearly ubiquitous expression of the *ACE2* and *FURIN* genes in the human body. Particularly dangerous seems the potential ability of the the SCARS-CoV-2 virus to infect the immune cells. Taken together with predominantly cell type-specific patterns of expression of genetic repressors and activators of the ACE2 and FURIN expression it may complicate the development of universally effective therapeutics. The availability of many genetically-relevant transgenic mouse models, in particular, the *Furin* null mice, should be regarded as a considerable advantage for preclinical development of drug candidates tailored to target the coronavirus infection. Specifically, the potential therapeutic utility of the highly selective ($K_{i,}$ 600 pm) intrinsically-specific FURIN inhibitor (a1-antitrypsin Portland (a1-PDX); Jean et al., 1998) should be tested in the immediate future.

**Methods**

**Data source and analytical protocols**

All data analyzed in this study were obtained from the publicly available sources. Gene set enrichment analyses (GSEA) were carried-out using the Enrichr bioinformatics platform, which enables the interrogation of nearly 200,000 gene sets from more than 100 gene set libraries. The Enrichr API (January 2020 through March 2020 releases) (Chen et al., 2013; Kuleshov et al., 2016) was used to test genes linked to the ACE2 and FURIN genes (or other genes of interest) for significant enrichment in numerous functional categories. In all tables and plots (unless stated otherwise), in addition to the nominal p values and adjusted p values, the "combined score" calculated by Enrichr is reported, which is a product of the significance estimate and the magnitude of enrichment (combined score $c = \log(p) * z$, where p is the Fisher's exact test p-value and z is the z-score deviation from the expected rank). Validation of the GSEA findings were carried-out employing the computational retrievals and manual curations of the gene expression profiles of the Gene Expression Omnibus (GEO) database.

**Statistical Analyses of the Publicly Available Datasets**

All statistical analyses of the publicly available genomic datasets, including error rate estimates, background and technical noise measurements and filtering, feature peak calling, feature selection, assignments of genomic coordinates to the corresponding builds of the reference human genome, and data visualization, were performed exactly as reported in the original publications (Glinsky, 2015-2020; Glinsky and Barakat, 2019; Glinsky et al., 2019; Guffanti et al., 2018) and associated references linked to the corresponding data visualization tracks (http://genome.ucsc.edu/). Any modifications or new elements of statistical analyses are described in the corresponding sections of the Results. Statistical significance of the Pearson correlation coefficients was determined using GraphPad Prism version 6.00 software. Both nominal and Bonferroni adjusted p values were estimated. The statistical significance between the mean values was estimated using the Student T-test. The significance of the differences in the numbers of events between the groups was calculated using two-sided Fisher's exact and Chi-square test, and the significance of the overlap between the events was determined using the hypergeometric distribution test (Tavazoie et al., 1999).

**Supplemental Information**

Supplemental information includes Supplemental Figures S1-S13. Supplemental information is available upon request.

**Author Contributions**

This is a single author contribution. All elements of this work, including the conception of ideas, formulation, and development of concepts, execution of experiments, analysis of data, and writing of the paper, were performed by the author.

**Acknowledgements**

This work was made possible by the open public access policies of major grant funding agencies and international genomic databases and the willingness of many investigators worldwide to share their primary research data. This work was supported in part by the OncoScar, Inc.